\documentclass[3p,times,procedia]{elsarticle}
\usepackage{nupha_ecrc}

\volume{00}

\firstpage{1}

\journalname{Nuclear Physics A}

\runauth{M.~Nahrgang et al.}

\jid{nupha}

\jnltitlelogo{Nuclear Physics A}

\usepackage{amssymb}

\usepackage[figuresright]{rotating}

\begin{document}

\begin{frontmatter}



%
\dochead{XXVIth International Conference on Ultrarelativistic Nucleus-Nucleus Collisions\\ (Quark Matter 2017)}
%

\title{Baryon number diffusion with critical fluctuations}


\author[1]{Marlene Nahrgang}
\author[2]{Marcus Bluhm}
\author[3]{Thomas Sch\"{a}fer}
\author[4]{Steffen A. Bass}

\address[1]{SUBATECH UMR 6457 (IMT Atlantique, Universit\'e de Nantes,
IN2P3/CNRS), 4 rue Alfred Kastler, 44307 Nantes, France }
\address[2]{Institute of Theoretical Physics, University of Wroc\l{}aw, 50204 Wroc\l{}aw, Poland}
\address[3]{Physics Department, North Carolina State University, Raleigh, NC 27695, USA}
\address[4]{Department of Physics, Duke University, Durham, NC 27708-0305, USA}

\begin{abstract}
The description of dynamical fluctuations near the QCD critical point in heavy-ion collisions is crucial for understanding 
the existing and upcoming experimental data from the beam energy scan programs. In this talk we discuss the evolution of 
fluctuations of the net-baryon density as given by a stochastic diffusion equation. We study equilibrium as well as dynamical 
systems for which we can show the impact of nonequilibrium effects on the second-order moment.
\end{abstract}


\end{frontmatter}


\section{Stochastic diffusion equation}
\label{sec:sde}
Near the conjectured critical point of QCD the net-baryon density $n_B$ becomes the critical mode, which exhibits diffusive 
dynamics \cite{Hohenberg:1977ym, Son:2004iv,Fujii:2004jt}. It is therefore expected that for beam energies for which 
the created matter reaches temperatures and baryochemical potentials in the critical region, fluctuations in the net-baryon 
number are enhanced \cite{Stephanov:1998dy,Stephanov:1999zu,Stephanov:2008qz,Asakawa:2009aj}. Due to the fast dynamics of 
the expanding strongly interacting matter nonequilibrium effects become important for a quantitative description of the 
critical fluctuations \cite{Berdnikov:1999ph,Nahrgang:2011mg,Kitazawa:2013bta,Mukherjee:2015swa}. Dynamical models of 
fluctuations coupled to fluid dynamical expansions have so far considered only fluctuations in the chiral order parameter, 
see N$\chi$FD \cite{Nahrgang:2011mg, Nahrgang:2011mv, Nahrgang:2011vn, Herold:2013bi,Herold:2016uvv}, which imprint traces 
in the cumulants of the event-by-event net-proton distributions \cite{Herold:2016uvv,Adamczyk:2013dal}. In this talk we 
study a numerical implementation of the stochastic diffusion equation for net-baryon number near the critical point
\begin{equation}
 \partial_t n_B(t,x) = \kappa \nabla^2\bigg(\frac{\delta {\cal F}[n_B]}{\delta n_B}\bigg) + \nabla J(t,x) \, .
 \label{eq:diffeq}
\end{equation}
Here, the free energy functional 
\begin{equation}
  {\cal F}[n_B] = T\int{\rm d}^3 x \,\left(\frac{m^2}{2n_c^2}\left(\Delta n_B\right)^2\right)
\end{equation}
for $\Delta n_B = n_B-n_c$, where $n_c$ is the critical net-baryon density, contains only a Gaussian term, such that 
Eq. (\ref{eq:diffeq}) is linear in $n_B$. 
Criticality is included in the mass $m$ via the temperature dependence of the equilibrium correlation length $\xi=\xi(T)$, 
where $m^2\propto1/\xi(T)^2$ as shown in Fig. \ref{fig:static} (left) is motivated by the 3d Ising model equation of 
state \cite{Bluhm:2016byc}.

The stochastic current $J$ describes the internal thermal fluctuations and is given by a Gaussian white noise variable 
$\zeta(t,x)$ and a variance determined from the fluctuation-dissipation theorem
\begin{equation}
 J(t,x) = \sqrt{2 T \kappa} \,\zeta(t,x)\, .
\end{equation}

The present system is a simple, linear version of fluid dynamical fluctuations capturing the essential critical phenomena up 
to second-order moments of the fluctuations. A full implementation of fluid dynamical fluctuations including fluctuations in 
the energy-momentum tensor and the coupling of Eq. (\ref{eq:diffeq}) to energy flow is a challenge for theory and numerical 
implementations \cite{Nahrgang:2017oqp}.

In the following we present results for static and fully equilibrated systems in section \ref{sec:eqflucs} and for systems 
with evolving temperature in section \ref{sec:dynflucs}.

\section{Static equilibrium fluctuations}
\label{sec:eqflucs}

\begin{figure}
 \includegraphics[width=0.455\textwidth]{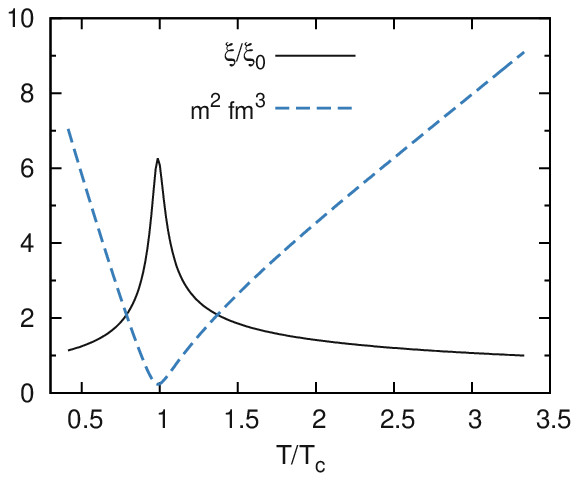}\hfill
 \includegraphics[width=0.48\textwidth]{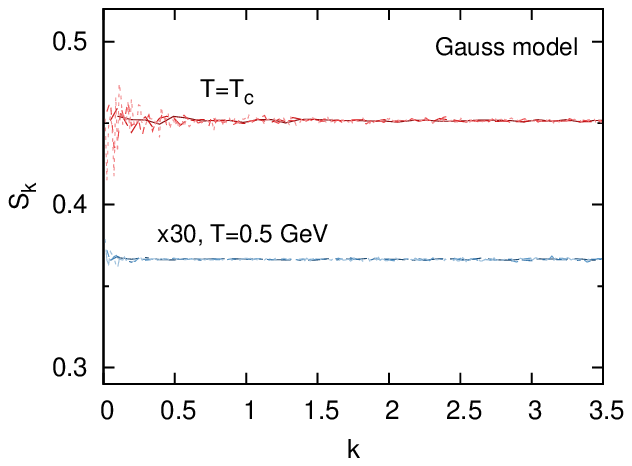}
 \caption{\label{fig:static}
 Left: Temperature dependence of $\xi$ and $m^2$ considered in this study, $T_c=0.15$~GeV and $\xi_0=0.48$~fm. 
 Right: Static structure factor $S_k$ as a function of momentum $k$ for $T=T_c$ and $T=0.5$~GeV for different 
 $\Delta x$ ($L=10$~fm and $N_x=64,\,128,\,256,\,512$). The fluctuations seen at small $k$ are of statistical origin. 
 }
\end{figure}

We solve Eq. (\ref{eq:diffeq}) for a one-dimensional system of length $L$ with periodic boundary conditions in an implicit 
scheme considering first a fixed diffusion coefficient $D=1$~fm expressed via $\kappa = D n_c/T$ and $n_c=1/(3$fm$^3)$. 
Equilibrium results are independent of the transport coefficient and solely depend on temperature and the lattice spacing 
$\Delta x=L/N_x$. The first quantity to verify our numerical algorithm against is the static structure factor 
$S_k=\langle\Delta n_B(k,0)\Delta n_B(-k,0)\rangle$. Numerical results for the discrete structure factor for two 
temperatures are shown in 
Fig.~\ref{fig:static} (right) for different lattice spacings. Within the statistical noise at small momenta, the 
theoretically expected perfectly flat behavior is reproduced. The implicit scheme applied here does not introduce any 
additional dependences on $\Delta x$ coming from the discretization.

\begin{figure}
 \includegraphics[width=0.48\textwidth]{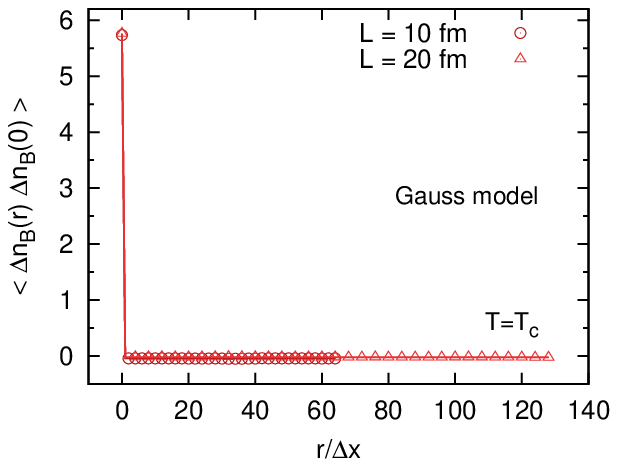}\hfill
 \includegraphics[width=0.48\textwidth]{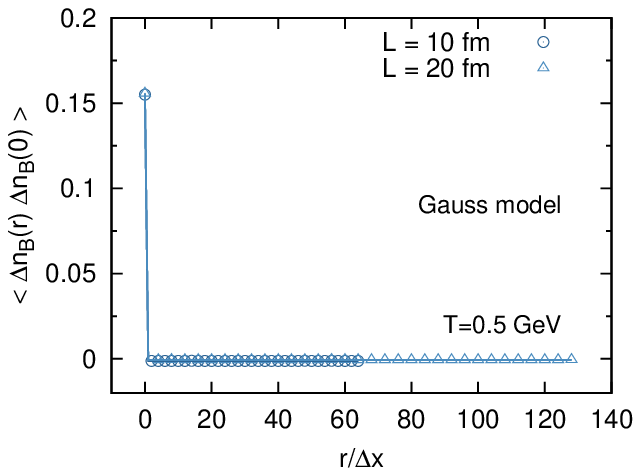}
 \caption{\label{fig:correl}
 Equal-time correlation function for $T=T_c$ and $T=0.5$~GeV as a function of distance $r$ in units of fixed $\Delta x=0.078125$~fm 
 for different $L$.}
\end{figure}

In Fig. \ref{fig:correl} we show the spatial correlation function 
$\langle\Delta n_B(r)\Delta n_B(0)\rangle=\int dk \,e^{ikr} S_k/(2\pi)$ of density fluctuations for 
$T=T_c$ (left) and $T=0.5$~GeV (right). We compare two different system sizes with the same resolution $\Delta x$. One observes 
that the fluctuations are uncorrelated over distances larger than $\Delta x$ and that the numerical results reproduce the analytical 
expectation $\langle\Delta n_B(r)\Delta n_B(0)\rangle = (n_c^2/m^2)\,\delta(r)$. Net-baryon number conservation in the 
finite-size system, which is satisfied in the numerics, results in a small but finite shift from this continuum 
expectation, $-n_c^2/(m^2N_x)$, which vanishes in the thermodynamic limit. At $r=0$, we obtain the local variance 
$\sigma^2$ of the density fluctuations, which is inversely proportional to the lattice spacing. Obviously, $\sigma^2$ is enhanced 
at the critical temperature compared to high temperatures. 

In Fig. \ref{fig:vartemp} (left) this variance is shown as a function of temperature. Again, the increase around $T_c$ is 
clearly visible and analytical expectations are perfectly reproduced. We also note that higher-order moments, like the kurtosis 
$\kappa$, are vanishing as expected for purely Gaussian models.

\section{Dynamical fluctuations}
\label{sec:dynflucs}
\begin{figure}
 \includegraphics[width=0.468\textwidth]{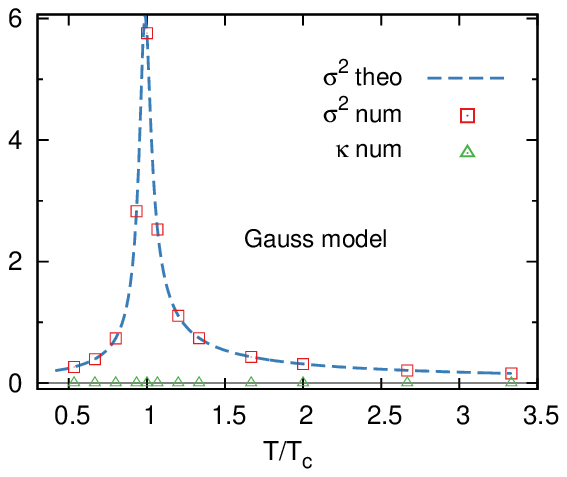}\hfill
 \includegraphics[width=0.48\textwidth]{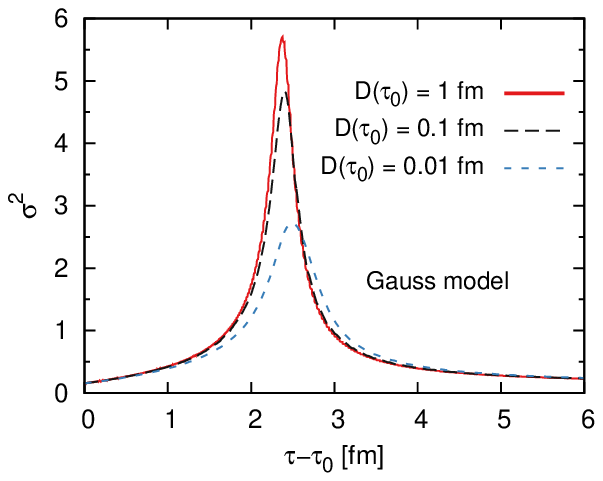}
 \caption{\label{fig:vartemp} Left: Static equilibrated fluctuations as a function of temperature from the numerical simulation 
 compared to theoretical expectations. Right: Dynamical fluctuations as a function of time.}
\end{figure}

Next, we turn to dynamical fluctuations by assuming a (spatially constant but) time-dependent temperature 
\begin{equation}
T(\tau)=T_0\left(\frac{\tau_0}{\tau}\right)^{dc_s^2}
\end{equation}
with $d=3$, $c_s^2=1/3$. The system is first equilibrated at high temperature $T_0=0.5$~GeV, before the cooling begins at 
$\tau_0=1$~fm/c. The critical temperature is reached at $\tau-\tau_0=2.33$~fm/c. In Fig. \ref{fig:vartemp} (right) we compare the 
results for three different initial diffusion coefficients $D(\tau_0)=1,0.1,0.01$~fm. For the 
fastest diffusion process the dynamical values of the variance come close to the equilibrium values around $T_c$. With decreasing 
diffusion strength nonequilibrium effects become stronger and lead to a decrease of the maximally achieved variance. A slight shift 
of the maximum to later times is also observed. This retardation effect is, however, rather weak in the present purely Gaussian model.

\section{Discussion}
We have benchmarked a numerical realization of the Gaussian stochastic diffusion equation versus analytical results for static and 
equilibrated systems. Because of the uncorrelated nature of the fluctuations in the net-baryon density, we find that the local 
variance depends on the lattice spacing. This unphysical situation will be cured in future work by the introduction of a surface 
tension term. We have observed an increase of the local variance near the critical temperature. We have additionally verified that 
higher-order moments, like the skewness and kurtosis, are vanishing, hence that propagated net-baryon number fluctuations are again 
purely Gaussian. Finally, we investigated systems in which the (spatially constant) temperature changes with time. In this case the 
dynamics is subject to nonequilibrium effects and becomes relaxational. We observe a reduction of the maximal variance and a shift 
of this maximum to later times corresponding to temperatures below $T_c$. Depending on the transport coefficient these nonequilibrium 
and retardation effects are more or less pronounced.

In ongoing work we include non-Gaussian couplings in the free energy functional motivated by the $3$d Ising universality class and 
investigate in particular the real-time dynamics of non-Gaussian cumulants \cite{tobepublishedsoon}.

\section*{Acknowledgments}
M.N. acknowledges support from the TOGETHER Project R\'{e}gion Pays de la Loire (France). The work of M.B. is funded by the European 
Union's Horizon~2020 research and innovation programme under the Marie Sk\l{}odowska Curie grant agreement No 665778 via the National 
Science Center, Poland, under grant Polonez UMO-2016/21/P/ST2/04035. This work was supported in parts by the U.S. Department of 
Energy under grants DE-FG02-03ER41260 and DE-FG02-05ER41367. The authors acknowledge fruitful discussions within the Beam Energy Scan 
Theory (BEST) Topical Collaboration. 




\end{document}